\documentclass[lettersize,journal]{IEEEtran}
\usepackage{amsmath,amsfonts}
\usepackage[ruled,vlined,linesnumbered]{algorithm2e}
\usepackage[noend]{algpseudocode}
\usepackage{array}
\usepackage[labelfont=scriptsize,textfont=scriptsize]{subfig}
\usepackage{textcomp}
\usepackage[numbers,sort&compress]{natbib}
\usepackage{enumitem}
\usepackage{stfloats}
\usepackage{tcolorbox}
\usepackage{url}
\usepackage{threeparttable}
\usepackage{algpseudocode}
\usepackage{verbatim}
\usepackage{graphicx}

\begin{document}
\title{Rethinking Membership Inference Attacks Against Transfer Learning}

\author{
	\small Cong Wu$^1$, Jing Chen$^2$$^{*}$, Qianru Fang$^2$, Kun He$^2$, Ziming Zhao$^3$, Hao Ren$^1$, Guowen Xu$^4$, Yang Liu$^1$, and Yang Xiang$^5$\\
	\emph{$^1$ Nanyang Technological University, Singapore.} 
	\emph{$^2$Wuhan University, China}
	\emph{$^3$Northeastern University, Boston, USA.}\\
	\emph{$^4$City University of Hong Kong, Hong Kong.}
	\emph{$^5$Swinburne University of Technology, Australia.}\\
 \texttt{\{cong.wu,hao.ren,yangliu,\}@ntu.edu.sg} \qquad
 \texttt{\{chenjing,hekun,fangqr\}@whu.edu.cn}\\
\texttt{guowenxu@cityu.edu.hk}\qquad
 \texttt{z.zhao@northeastern.edu}\qquad
\texttt{yxiang@swin.edu.au}
 \thanks{$^{*}$ Corresponding author: Jing Chen.}
}
\maketitle

\begin{abstract}
	Transfer learning, successful in knowledge translation across related tasks, faces a substantial privacy threat from membership inference attacks (MIAs). These attacks, despite posing significant risk to ML model's training data, remain limited-explored in transfer learning.
The interaction between teacher and student models in transfer learning has not been thoroughly explored in MIAs, potentially resulting in an under-examined aspect of privacy vulnerabilities within transfer learning.
	 In this paper, we propose a new MIA vector against transfer learning, to determine whether a specific data point was used to train the teacher model while only accessing the student model in a white-box setting.
	 Our method delves into the intricate relationship between teacher and student models, analyzing the discrepancies in hidden layer representations between the student model and its shadow counterpart. These identified differences are then adeptly utilized to refine the shadow model's training process and to inform membership inference decisions effectively.
	 Our method, evaluated across four datasets in diverse transfer learning tasks, 
	 reveals that even when an attacker only has access to the student model, the teacher model's training data remains susceptible to MIAs. 
We believe our work unveils the unexplored risk of membership inference in transfer learning.
\end{abstract}

\section{Introduction}

Transfer learning has witnessed a marked increase in adoption recently, primarily due to its proficiency in leveraging knowledge from one domain to enhance performance in another, closely related domain~\cite{wang2018great,zhu2023transfer,fang2024ic3m}. This method enables the application of complex models developed with extensive data~\cite{lin2024efficient,liang2024towards,yuan2025constructing,yuan2023graph,lin2022channel}, such as those by large corporations or hospitals, to be adapted for use in smaller entities like startups or smaller hospitals, thereby bypassing the need for substantial labeled data. However, this widespread adoption also raises concerns regarding the potential for privacy breaches, particularly through MIAs~\cite{shokri2017membership,lin2024fedsn,zhang2024generated,lin2024hierarchical,li2020label,nasr2018machine,zhang2024satfed,hu2023source}, as sensitive information, including biometric and healthcare data, is increasingly shared among different organizations~\cite{jignesh2020face}. Consequently, while transfer learning offers numerous advantages, such as customizing models to specific tasks without extensive data, it also necessitates stringent scrutiny to prevent potential privacy violations.

In MIAs, adversaries exploit access to a model's parameters or its output to ascertain if a specific data point was used during the model's training~\cite{shokri2017membership,salem2018ml}. They typically achieve this by analyzing how the model behaves with a given data point compared to others, potentially identifying it as part of the training set based on output discrepancies or changes when the point is included or excluded from training data. 
The rising trend of data sharing, particularly sensitive personal information, among organizations underscores the critical need for stringent privacy measures, guided by regulations such as the General Data Protection Regulation (GDPR) \cite{gdpr} and the California Consumer Privacy Act (CCPA) \cite{ccpa}. These regulations are pivotal in upholding data privacy, especially within transfer learning where knowledge transfer occurs across varied organizational echelons~\cite{wuTouch2024,Maria2024,wang2019differentially,gao2019privacy}.

In transfer learning, research efforts like those from Zou et al. and Hidano et al.~\cite{zou2020privacy,hidano2021transmia} have focused on black-box MIAs. Zou et al.~\cite{zou2020privacy} were pioneers in this field, applying shadow training to analyze MIAs against the target model, albeit without extending this analysis to the teacher model. Hidano et al.~\cite{hidano2021transmia} advanced this research by introducing transfer shadow training, enhancing attack accuracy by exploiting the transferred model's parameters. However, neither study fully explored the critical relationship between teacher and student models in a white-box way, a key aspect for a comprehensive understanding of MIAs in transfer learning. Recognizing and exploring this relationship is crucial for a deeper and more accurate assessment of privacy risks in this domain.
Thus, we are motivated by the question of how a white-box attack framework, attentively analyzing the complex interactions between teacher and student models, can yield a more comprehensive and nuanced understanding of privacy risks in transfer learning environments.

\textbf{Our approach.} We propose a new MIA vector in transfer learning to determine if a data point was part of the teacher model's training by analyzing the student model. In our \emph{white-box} scenario, the attacker has full access to the student model's internal architecture and feature representations. Although the attacker does not have direct access to the teacher model or its data, they use a shadow dataset similar to the student model's training data, classifying it into member and non-member groups to train a shadow student model~\cite{salem2018ml,long2017towards,song2019privacy,shokri2017membership}. The attacker then queries the actual and shadow student models using the shadow dataset to collect feature representations and analyze variations.

The attacker establishes adaptive thresholds based on these feature variations to infer membership. This strategic method exploits the differences between the actual and shadow student models' feature representations, revealing patterns indicative of membership in the teacher model's training data. By using unique data transformation discrepancies in transfer learning, our approach significantly enhances the efficacy of MIAs. This highlights privacy risks in transfer learning, especially given the increasing reliance on private or semi-private datasets, emphasizing the need for stronger privacy safeguards.

Significantly different from prior MIA researches~\cite{zou2020privacy,hidano2021transmia,ying2020privacy,liew2020faceleaks,jia2019memguard,shokri2017membership,nasr2018machine}, we unveil a novel white-box MIA vector specifically tailored for transfer learning contexts, addressing a gap not extensively covered in existing literature. 
This attack distinctively analyzes differential representations between teacher and student models,
a step beyond the traditional focus on direct feature comparisons, thus providing a nuanced understanding of transfer-induced vulnerabilities. We also innovate with a trinary decision framework, which refines the attack precision by distinguishing among various membership statuses more accurately. Our investigation into the complex interplay between teacher and student models uncovers subtle yet critical insights into their shared vulnerabilities, advocating for enhanced privacy measures in transfer learning. Through a meticulous comparative analysis using shadow models, our work underscores the need for a reevaluated defense strategy, positioning our contributions as a pivotal reference for future research in securing transfer learning mechanisms.

\textbf{Contribution.}  We summarize our contribution as follows:
\begin{itemize} [leftmargin=*]
	\item We unveil a new MIA vector in transfer learning, which is the first white-box attack against transfer learning to the best of our knowledge.
	It employs differential representation analysis and a ternary decision framework to elucidate privacy vulnerabilities in teacher-student model interactions.  

	\item To elucidate the interplay between teacher and student models, we present an adaptive threshold selection mechanism, enabling the extraction of representation differences between the student model and corresponding shadow model (\S\ref{sec: attackarchitecture}). 

	\item Our approach is rigorously tested across diverse settings, four datasets, and different transfer learning tasks, comparing it against current state-of-the-art methods. The results indicate that our proposed attack achieves a high attack accuracy, and more effectively exposes the teacher model's membership privacy then previous black-box MIAs (\S\ref{sec:exp}). 
 
	\item We systematically consider other two possible attacks under transfer learning, 
	including inferring the student membership using the knowledge of the student model, 
	and inferring the teacher membership using the knowledge of the teacher model.
	We also evaluate the performance of these attacks cases and compare them with our attack cases (\S\ref{sec:othertwo}). 
\end{itemize}

\section{Background}
\label{sec:Background}
This section briefs transfer learning and MIA.

\subsection{Transfer Learning}

Transfer learning leverages a pre-trained teacher model's architectural framework and layer weights to enhance a student model's efficacy on related tasks~\cite{lu2022reinforcement,wu2022echohand}. Its proven effectiveness is particularly beneficial for entities with limited data or computational power, enabling the development of precise models with optimized resource utilization for task-specific applications.

\begin{figure}[!t]
	\centering
	\includegraphics[width = 0.55\linewidth]{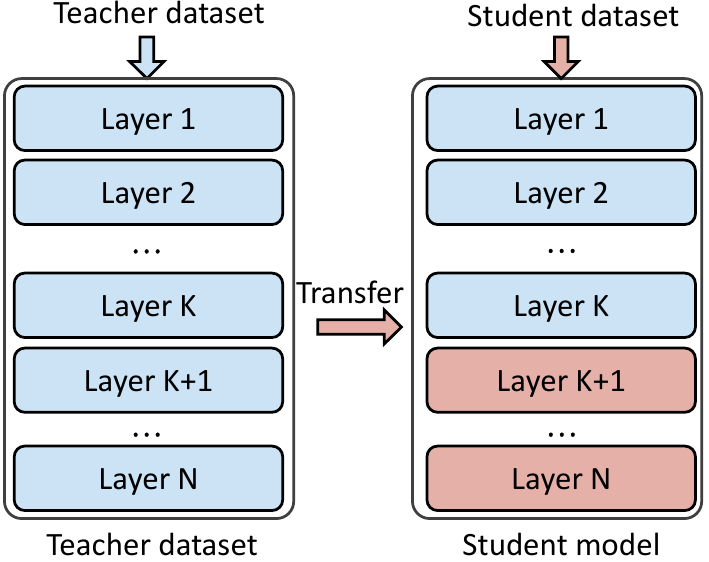}
	\caption{Illustration of transfer learning} 
	\label{fig:tl} 
\end{figure}

As illustrated in Figure~\ref{fig:tl}, transfer learning begins with the student model adopting the teacher model's feature extraction layers and appending a new dense layer to fit its specific task. 
In subsequent training phases, the student model, using a smaller dataset, retains fixed weights in the initial K feature extraction layers, while adapting the weights in the remaining layers. These extraction layers are crucial for recognizing input data patterns through convolution and pooling, while the dense layers synthesize this information for specific outputs or classifications. Maintaining the pre-trained layers unchanged capitalizes on their learned features, optimizing training efficiency and reducing resource demands. The extent of layer freezing, determined by the similarity between the teacher's and student's tasks, varies from minimal adjustments for closely related tasks to more extensive tuning for distinct ones, ensuring the student model's relevancy and performance.

\subsection{Membership Inference Attack}

MIA aims to ascertain if a deep learning model, often involving complex nonlinear functions and numerous layers, was trained utilizing a given data record~\cite{carlini2022membership,shokri2017membership}. This process of inquiry requires the attacker to interact with the model, supplying the given data point and observing the output to assess the membership status of the provided data. 
This assessment, a key facet of ML privacy research, can be formally represented as in Eq.~\ref{eq:mia}. 
The primary attack task is to construct a MIA model $\mathcal{A}$ that operates against target model $\mathcal{M}$, with the aim to predict the membership status of a given data point $\mathbf{x}_{target}$.
\begin{align}
\label{eq:mia}
\footnotesize	
\mathcal{A}(\mathbf{x}_{\text{target}}, \mathcal{M}, \Omega) = 
\begin{cases} 
	1, & \text{if } \mathbf{x}_{\text{target}} \text{ is in the training data of } \mathcal{M} \\
	0, & \text{otherwise}
\end{cases}
\end{align}
The target model $\mathcal{M}$, a pre-trained and possibly heavily parametrized model, and the prior knowledge $\Omega$, often encapsulating assumptions regarding the data distribution or the target model's characteristics, can be accessed by the attacker. The attack model $\mathcal{A}$ serves as a binary classifier, tasked with determining whether the given data point originated from the training set. It outputs the indicative of the fact that the input data record $\mathbf{x}_{target}$ is a constituent of the target model $\mathcal{M}$'s training dataset.

In terms of the level of insight an attacker has into the target model and its training data, we can distinguish between two forms of MIAs: black-box and white-box. In the scenario of a black-box attack, the attacker's access is limited to output of target model. This permits them to extract the prediction vectors, namely the softmax output, but their insight ends here.
Shifting to the white-box attack, the situation is vastly different. The attacker enjoys extensive access, stretching to the complete details of the model: the weights, architecture, and at times, even the gradients. This increased accessibility endows the attacker with a more holistic comprehension of the functioning of the model.
To ensure a smooth transition between these two scenarios, it's vital to highlight that while the black-box approach restricts the attacker's knowledge, the white-box approach amplifies it, encompassing even the minutest of model details. Thus, the threat level in a white-box attack escalates due to the comprehensive understanding it affords the attacker.

\section{Attack Overview}
\label{sec:attackoverview}
In this section, we present the attack scenario, then illustrate threat model and problem formulation.

\subsection{Attack Scenario} 
In practical scenarios, base models are often not trained solely on publicly available datasets like ImageNet. Instead, they might incorporate proprietary data to cater to specialized and sensitive applications like fraud detection using financial records~\cite{patel2023credit,alsuwailem2023performance}, disease diagnosis with personalized medical imaging~\cite{lin2024fedlppa,liu2024efficient}, or biometric recognition in private systems~\cite{sitova2015hmog,cariello2024smartcope,cardaioli2020your}. 
Commercialized models often include meta-information about their training context that attackers can exploit. Platforms like Hugging Face\footnote{\url{https://huggingface.co/models}}, 
Azure AI Services\footnote{\url{https://azure.microsoft.com/en-us/products/ai-services/}}, 
TensorFlow Hub\footnote{\url{https://www.tensorflow.org/hub}} frequently provide dataset descriptions (e.g., ImageNet, COCO), training methodologies, hyperparameters, benchmarks, and README files outlining training purposes and fine-tuning strategies. This detailed information allows attackers to replicate the data distribution and architecture in shadow models, facilitating realistic MIAs. For instance, if a model is known to have been trained on a proprietary medical imaging dataset for disease diagnosis, attackers can simulate a similar dataset and shadow model, enabling them to compare feature representations and infer patterns indicative of the original training data. This comprehensive level of detail makes shadow model creation straightforward and heightens the risks associated with MIAs.

In the context of transfer learning, an attacker aims to determine whether specific data were part of the teacher model's training by analyzing the intermediate outputs of the student model. Despite being trained on a distinct, often smaller, dataset, the student model retains information from the teacher model through the transfer learning process. Thus, having access to the student's outputs offers clues that can reveal the original training data.

The attacker systematically queries the student model, scrutinizing its intermediate outputs to detect patterns indicative of the teacher model's training data. They can construct shadow models mimicking the student model's behavior to analyze discrepancies in feature representations. Comparing shadow and student models, the attacker identifies discrepancies and infers the likelihood of data belonging to the teacher model's training set. This approach is practical where adversaries can observe or intercept model outputs, such as in shared or distributed computing environments. Gaining insights into intermediate outputs uncovers vulnerabilities in transfer learning, emphasizing the importance of securing these outputs.

\subsection{Threat Model}

Our research expands upon this by exploring the white-box attack scenario, where the attacker can access model's intermediate output. Specifically, we consider the following attack cases that are possible in  transfer learning: 
MIA against the teacher model while accessing the student model (At.T \& Ac.S), 
directly MIA against the accessible teacher model (At.T \& Ac.T), 
and directly MIA against the accessible student model (At.S \& Ac.S), thereby providing a comprehensive investigation into the vulnerabilities across transfer learning frameworks.

\begin{itemize} [leftmargin=*]
\item \textbf{At.T \& Ac.S.}
The attacker has prior knowledge of the student model's parameters, structure, parameters, and training data distribution, and also has access to the student model itself. 
The goal is to decide if a particular data point was used in the training process of the teacher model, without having any knowledge of the teacher model. To achieve this, the attacker may utilize the shadow model dataset and the student model to generate feature sets and adaptive thresholds for MIAs.

\item \textbf{At.T \& Ac.T.}
The attacker has none knowledge of the student model, but has prior knowledge of the teacher model. 
The attacker's goal is to decide if a specific data record was used to train the teacher model.
	
\item \textbf{At.S \& Ac.S.}
The attacker has prior knowledge of the student model but lacks any knowledge about the teacher model. 
The attacker attempts to perform MIA toward the student model, i.e., deciding if a given input data record was used to train the student model.

\end{itemize}

Despite their similar attack pipeline, the complexity of At.T \& Ac.T and At.S \& Ac.S attack cases, 
differs significantly due to the varying sizes of training data used to train the teacher and student models. 
The teacher model is typically trained from scratch using a substantial amount of data, 
while the student model leverages a comparably smaller dataset for its training, achieved through fine-tuning the teacher model. As such, it is impractical to infer the student model's membership using solely the information from the teacher model, as the dataset used for the student model's training is not part of the teacher model's training process. In this paper, we aim to concentrate on these three potential attack cases, exploring the viability of MIAs in transfer learning.

We also note that our white-box assumption models an adversary with significant, yet plausible, system penetration capabilities, aligning with scenarios where internal threats or security breaches provide access to the model's internals, a consideration crucial for rigorous security evaluations.
The white-box access in our threat model is an intentional design to test the upper bounds of an attacker's capability, offering a new stringent MIA security evaluation framework; this higher threat level assumption is reasonable as it represents the full exploitation of an attacker's ability to pose serious security threats, ensuring preparedness for worst-case scenarios.
In real-world scenarios, attackers can also pragmatically deduce the nature of a base model's tasks by examining the student model's attributes, supported by associated metadata or domain-specific commonalities. This information enables them to craft shadow models that reflect the teacher's training context, making it a practical assumption for understanding and mitigating potential vulnerabilities in transfer learning frameworks.

\begin{table}
	\centering
	\caption{Comparison of three different attack cases}
	\label{tab:parameter}
	\resizebox{\linewidth}{!}{\begin{tabular}{cccc}
		\hline
		\hline
		Attack type   & Target & Knowledge & Formulation                                                                             \\
		\hline
		\hline
		At.T \& Ac.S  & Teacher & Student          & $\mathcal{A}_{At.T \& Ac.S}(\mathcal{M}_s(x_{target}), \Omega_s) \rightarrow \{0,1,2\}$ \\
		At.S \& Ac.S & Student & Student          & $\mathcal{A}_{At.S \& Ac.S}(\mathcal{M}_s(x_{target}), \Omega_s) \rightarrow \{0,1\}$  \\
		At.T \& Ac.T  & Teacher & Teacher          & $\mathcal{A}_{At.T \& Ac.T}(\mathcal{M}_t(x_{target}), \Omega_t) \rightarrow \{0,1\}$     \\
		\hline
	\end{tabular}}
\end{table}

\subsection{Attack Formulation}
\label{sec:problem}

The notions are given as Table~\ref{tab:notions}.
The formation of the three attack cases are given as follows.

\begin{table}
	\centering
	\scriptsize
	\caption{Notations in this paper}
	\label{tab:notions}
	\begin{tabular}{lc}
		\hline \hline
		Notation                         & Description                                      \\
		\hline \hline
		At.T \& Ac.S                     & Attack teacher model while accessing student model \\
		At.T \& Ac.T                     & Attack teacher model while accessing teacher model \\
		At.S \& Ac.S                     & Attack student model while accessing student model \\
		$\mathbf{x}_{target}$            & Target data point                              \\
		$\mathbf{x}_{train}$             & Training data point                            \\
		$\mathcal{M}_t$                  & Teacher model                                    \\
		$\mathcal{M}_s$                  & Student model                                    \\
		$\mathbf{D}^{shadow}_s$          & Shadow student dataset                           \\
		$\mathbf{D}^{shadow}_t$          & Shadow teacher dataset                           \\
		$\mathbf{D}^n$                   & Random generated non-membership dataset          \\
		$\mathcal{M}_t'$                 & Shadow teacher model                             \\
		$\mathcal{M}_s'$                 & Shadow student model                             \\
		$D_{train}$                      & Training dataset                                 \\
		$\sigma_1$, $\sigma_2$, $\sigma_3$ & The three selected thresholds                    \\
		$\mathcal{A}$                    & The attack model                                 \\
		$\Omega_s$                       & The prior knowledge from the student model       \\
		$\Omega_t$                       & The prior knowledge from the teacher model       \\
		${}^m\mathbf{D}^{shadow}_s$      & The member data of shadow student model          \\
		${}^n\mathbf{D}^{shadow}_s$      & The non-member data of shadow student model      \\
		${}^m\mathbf{D}^{shadow}_t$      & The member data of shadow teacher model          \\
		${}^n\mathbf{D}^{shadow}_t$      & The non-member data of shadow teacher model      \\
		\hline
	\end{tabular}
\end{table}

\textbf{At.T \& Ac.S.}
We formulate the problem of At.T \& Ac.S as Eq.~\ref{eq:attacs}:
\begin{equation}
\label{eq:attacs}
\mathcal{A}_{At.T \& Ac.S}(\mathcal{M}_s(\ \mathbf{x}_{target}), \Omega_s) \rightarrow \{0,1,2\}
\end{equation}
where $\mathcal{A}_{At.T \& Ac.S}$ is the attack model.
$\mathcal{M}_s$ is the student model.
$\mathcal{M}_s( \mathbf{x}_{target})$ is the output of student model.
where the attack model $A_{At.T \& Ac.S}$ is a three-class classifier.
$\Omega_s$ is the auxiliary knowledge from the student model, typically encompassing the model architecture, hyperparameters, or output behaviors observed during interactions with $M_s$.
Attackers can deduce such information through public documentation, direct model interaction, or inferential analysis based on common practices within the domain of the student model's application.  
1 represents that the data point is a member of $\mathcal{M}_t$'s training dataset.
2 represents that the data point is a member of $\mathcal{M}_s$'s training dataset.
0 represents that the data point is a non-member of $\mathcal{M}_t$'s training dataset and $\mathcal{M}_s$'s training dataset.

The intuition behind this attack stems from the premise that feature representations differ between a model trained on specific data and an analogous model trained independently. In transfer learning, these disparities in feature representations between the student and a shadow student model, especially when the latter is trained on disparate data, can reveal traces of the teacher model's training dataset. This approach capitalizes on the fundamental principle that learned representations in neural networks are inherently data-dependent, thereby justifying the use of feature representation differences as a mechanism for inferring membership in the teacher model's dataset.

\textbf{At.T \& Ac.T.}
We formulate At.T \& Ac.T problem as Eq.\ref{eq:attact}:
\begin{equation}
\label{eq:attact}
\mathcal{A}_{At.T \& Ac.T}(\mathcal{M}_t(\mathbf{x}_{target}), \Omega_t) \rightarrow \{0,1\}
\end{equation}
where the attack model $\mathcal{A}_{At.T \& Ac.T}$ is a binary classifier.
$\mathcal{M}_t$ is the teacher model.
$\mathbf{x}_{target}$ is the data used for membership inference.
$\mathcal{M}_t(\mathbf{x}_{target})$ is the output of teacher model, and
$\Omega_t$ is the auxiliary knowledge from the teacher model.
1 represents that the data point is a member of $\mathcal{M}_t$'s training dataset and 0 otherwise.

\textbf{At.S \& Ac.S.}
Given that the student model is trained based on the teacher model, 
a portion of the structure and parameters of the model are transferred to the student model. As a result, the privacy related to membership is encoded within the teacher model's parameters. Despite the retraining of the student model, the privacy implications persist. We represent the problem of At.S \& Ac.S as denoted in Eq.~\ref{eq:atsacs}.

\begin{equation}
\label{eq:atsacs}
\mathcal{A}_{At.S \& Ac.S}(\mathcal{M}_s(\mathbf{x}_{target}), \Omega_s) \rightarrow \{0,1\}
\end{equation}
where the attack model $\mathcal{A}_{At.S \& Ac.S}$ is a binary classifier.
0 represents that the data point is not a member of $\mathcal{M}_s$'s training dataset.
1 represents that the data point is a member of $\mathcal{M}_s$'s training dataset.

\begin{figure}[!t]
	\centering
	\includegraphics[width = \linewidth]{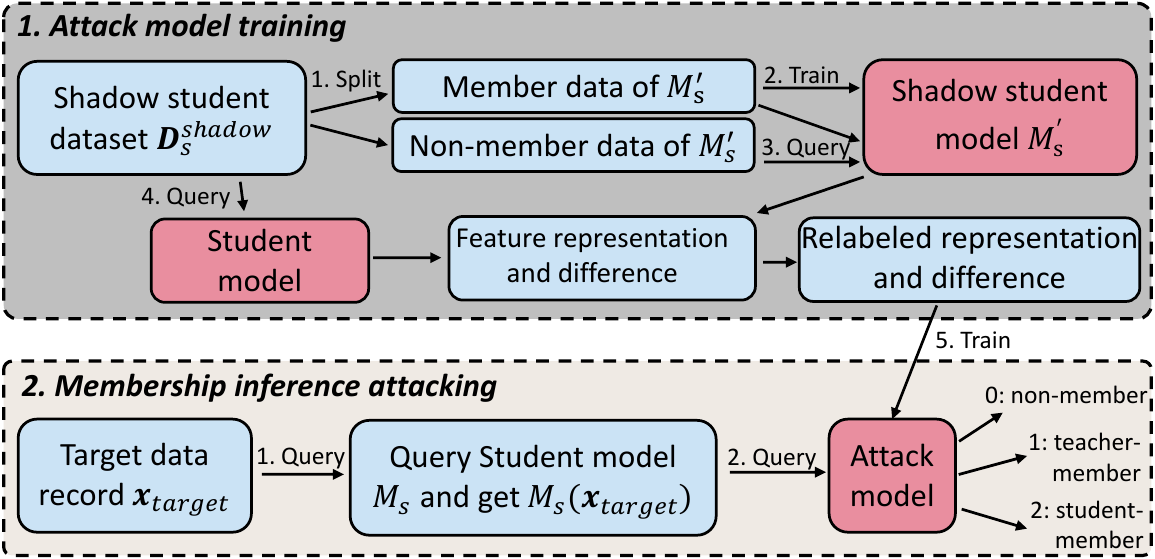}
	\caption{Attack workflow of At.T \& Ac.S.}
	\label{fig: attacs}
\end{figure}

\begin{figure*}[!t]
	\centering
	\captionsetup[subfloat]{labelfont=scriptsize,textfont=scriptsize}
	\subfloat[At.T \& Ac.T]{\includegraphics[width = 0.4\linewidth]{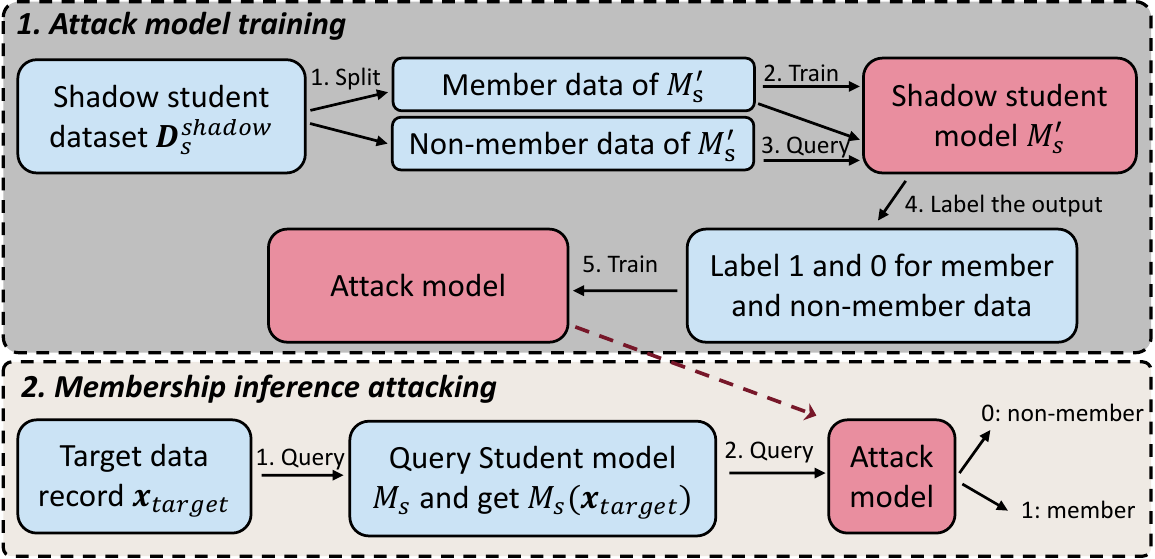}} \hspace{8mm}
	\subfloat[At.S \& Ac.S]{\includegraphics[width = 0.4\linewidth]{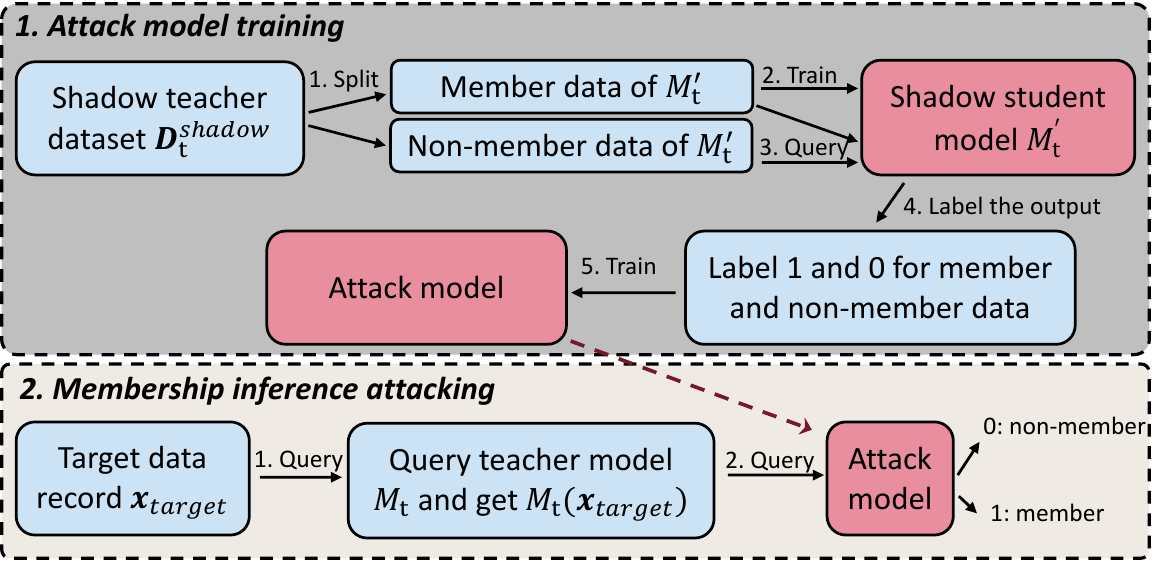}} \\
	\caption{Attack workflow of At.T \& Ac.T (a) and At.S \& Ac.S (b)}
	\label{fig:ovewflow2}
\end{figure*}

\section{Attack Design}
\label{sec: attackarchitecture}
In this section, we illustrate attack schemes of At.T \& Ac.S.

\subsection{Attack Model Training}
As shown in Figure~\ref{fig: attacs},
the attacker first splits the shadow student dataset into member and non-member datasets,
i.e., ${}^m\mathbf{D}_s^{shadow}$ and ${}^n\mathbf{D}_s^{shadow}$.
The shadow student model $\mathcal{M}_s'$ is trained using the member dataset.
Then the attacker queries the shadow student model and real student model using the member dataset respectively, and obtains the feature representation of the intermediate hidden layer, i.e.,
$\mathcal{M}_s'(\mathbf{x})$ and $\mathcal{M}_s(\mathbf{x})$, $\mathbf{x}\in {}^m\mathbf{D}_s^{shadow}$.
Then the attacker calculates the $L_2$ distance between the two feature representations,
i.e., $|\mathcal{M}_s'(\mathbf{x}) -  \mathcal{M}_s(\mathbf{x})|$, $\mathbf{x} \in {}^m\mathbf{D}_s^{shadow}$,
which represents the privacy of teacher dataset.

Typically, the tasks of the teacher and student models share substantial similarities, making the inference of the teacher model's membership challenging once the student model is trained based on the teacher model.
For instance, the teacher model might classify images of cats and dogs, while the student model could be tasked with classifying different breeds of cats. In such scenarios, it is considerably tough for an attacker to infer the teacher model's membership when querying the model with non-membership data.

The attacker also generates random noisy images, $\mathbf{D}^n$ where the generated images are without semantics.
The attacker queries the student model using the generated images and gets the feature representation of the intermediate layer $\mathcal{M}_s(\mathbf{x}), \mathbf{x} \in  \mathbf{D}^n$.
The feature representation of the intermediate hidden layer is then labeled with different labels.
Specifically,
the output feature representation queried using the member dataset of the shadow student model is labeled with 2, i.e.,
$(\mathcal{M}_s(\mathbf{x}),2), \mathbf{x}\in {}^m\mathbf{D}_s^{shadow} $ is labeled with 2.
The difference is calculated as the privacy of the teacher model member data, and is labeled with 1,
i.e., $(|\mathcal{M}_s'(\mathbf{x}) - \mathcal{M}_s(\mathbf{x})|,1), \mathbf{x}\in {}^m\mathbf{D}_s^{shadow}$.
Similarly, the student model is queried using the randomly generated images, and the output feature representation of the intermediate layer is labeled with 0,
i.e., $(\mathcal{M}_s(\mathbf{x}),0), \mathbf{x}\in \mathbf{D}^n$.
The attacker then uses the labeled data to train the membership inference model.
The model is trained to distinguish three different classes, i.e., teacher-member, student-member, and non-member.

\begin{figure}[!t]
	\centering
	\captionsetup[subfloat]{labelfont=scriptsize,textfont=scriptsize}
	\subfloat[]{\includegraphics[width=0.2\linewidth]{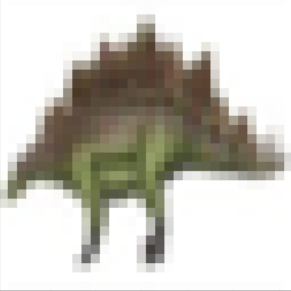}}
	\hspace{2mm}
	\subfloat[]{\includegraphics[width=0.2\linewidth]{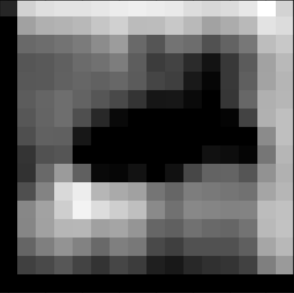}}
	\hspace{2mm}
	\subfloat[]{\includegraphics[width=0.2\linewidth]{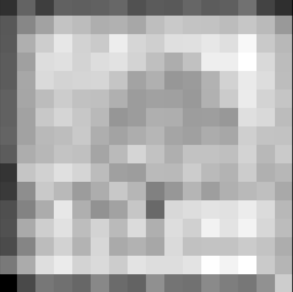}}
	\hspace{2mm}
	\subfloat[]{\includegraphics[width=0.2 \linewidth]{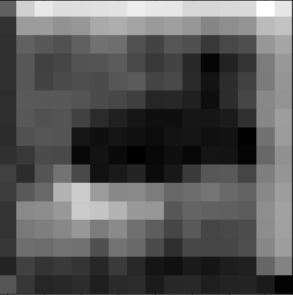}}\\
	\caption{A example of the raw image (a), the feature representation of the student model (b), shadow student model (c),  and feature representation differences (d)}
	\label{fig:example}
\end{figure}

Algorithm~\ref{alg:alg2} details the attack model training process.
The shadow student dataset \( \mathbf{D}_{s}^{shadow} \) is divided into the member dataset \( ^m\mathbf{D}_{s}^{shadow} \) and the non-member dataset \( ^n\mathbf{D}_{s}^{shadow} \) using a 1:1 ratio; specifically, 50\% of \( \mathbf{D}_{s}^{shadow} \) is randomly selected to train the shadow model, forming \( ^m\mathbf{D}_{s}^{shadow} \), while the remaining 50\% constitutes \( ^nD_{s}^{shadow} \), representing data that the shadow model has not encountered.
The threshold $\sigma_1$ critically impacts the classification accuracy by dictating the distinction between member and non-member data.
An optimal $\sigma_1$ minimizes both false positives and negatives, thereby enhancing the precision and recall of the attack model $\mathcal{A}$.
As an example, Figure~\ref{fig:example} present the feature representation of the student model, the shadow student model, and the feature representation differences.
The feature representations and the difference is then used to decide the membership of the teacher and student models.

\begin{algorithm}[!t]
	\SetAlgoLined
	\scriptsize
	\DontPrintSemicolon
	\KwIn{$\mathbf{D}_s^{shadow}$, $\mathbf{D}^n $, $\mathcal{M}_s$}
	\KwOut{The attack model $\mathcal{A}$}
	Split $\mathbf{D}_s^{shadow}$ into ${}^m\mathbf{D}_s^{shadow}$ and ${}^n\mathbf{D}_s^{shadow}$\\
	Train the shadow model $\mathcal{M}_s'$ using ${}^m\mathbf{D}_s^{shadow}$ \\
	Compile the teacher-member dataset $(|\mathcal{M}_s(\mathbf{x})-\mathcal{M}_s'(\mathbf{x})|,1 )$, $\mathbf{x} \in {}^m\mathbf{D}_s^{shadow} $ \\
	Compile the student-member dataset $(\mathcal{M}_s(\mathbf{x}),2 )$, $\mathbf{x} \in {}^m\mathbf{D}_s^{shadow} $ \\
	Compile the non-member dataset $(\mathcal{M}_s(\mathbf{x}),0	 )$, $\mathbf{x} \in \mathbf{D}^n $ \\
	Train the attack model $\mathcal{A}$ \\
	\Return{ $\mathcal{A}$ }
	\caption{Training attack model}
	\label{alg:alg2}
\end{algorithm}

\subsection{Attack Strategy}
During attack phase, the attacker queries the student model using the target data record $\mathbf{x}_{target}$,
and derives the feature representation of the intermediate hidden layer.
Then, the attacker queries the attack model using the output feature representation and
calculates the distance between the feature representation and the labeled data in the training set.
$L_2$ distance is used as the distance metric.
Specifically, there are three cases:
\begin{itemize}[leftmargin=*]
	\item  (i) $dis(\mathcal{M}_s(\mathbf{x}_{target}), \mathcal{M}_s(\mathbf{x})  )\textless \sigma_1$, $\mathbf{x}\in \mathbf{D}^{shadow}_s$.
	      This indicates the feature representation of the target data record is similar to  from the feature representation of the student shadow model member data.
	      The target can be regarded as a student-member data point and the attack result is 2.

	\item	 (ii)  $\sigma_1 \textless  dis(\mathcal{M}_s(\mathbf{x}_{target}), \mathcal{M}_s(\mathbf{x}))   \textless \sigma_2, \mathbf{x}\in \mathbf{D}^{shadow}_s$.
	      This indicates that the feature representation of the target data record is not the number of student number.
	      But it requires to further decide whether the target data record is a teacher-member data point.
	     The attack model calculates the $L_2$ distance between the feature representation and the labeled teacher-member data in the training set.
	      There are two special cases.
	      If $dis(\mathcal{M}_s(\mathbf{x}_{target}), |\mathcal{M}_s(\mathbf{x})-\mathcal{M}_s'(\mathbf{x})|),\mathbf{x}\in  \mathbf{D}^{shadow}_s \textless \sigma_3$. The target data record can be regarded as a teacher-member data point. The inference attack results is 1.
	      If $dis(\mathcal{M}_s(\mathbf{x}_{target}), |\mathcal{M}_s(\mathbf{x})-\mathcal{M}_s'(\mathbf{x})|),\mathbf{x}\in  \mathbf{D}^{shadow}_s \textgreater \sigma_3$. The target data record can be regarded as a non-member data point. The inference attack results is 0.

	\item 	  (iii) $dis(\mathcal{M}_s(\mathbf{x}_{target}), \mathcal{M}_s(\mathbf{x})) \textgreater  \sigma_2,\mathbf{x}\in \mathbf{D}^{shadow}_s$.
	      This indicates the feature representation of the target data record is different from the feature representation of the shadow model member data.
	      The inference attack results is 1.
\end{itemize}

In our approach, we use a distance metric $dis(\mathcal{M}_s(\mathbf{x}_{target}), \mathcal{M}_s(\mathbf{x}))$ to measure how closely a target sample's output from the student model aligns with known samples. This measurement is crucial for determining whether a sample is likely part of the model's training set. We set thresholds to categorize these distances, aiding in accurately distinguishing between members and non-members. These thresholds are carefully determined to ensure our attack can effectively identify membership status, demonstrating the practicality and ingenuity of our method within transfer learning's unique framework.

\subsection{Threshold Selection}
In the At.T \& Ac.S scenario for conducting MIAs, we establish three distinct thresholds to accurately determine if a data record was part of the training set. Our threshold selection algorithm outlines this process meticulously. For $\sigma_1$, we process the student shadow dataset through the student model, mirroring the student model's data distribution. We then calculate the distance between feature representations from the shadow student model and the actual student model data. The median value of these distances is chosen as the threshold $\sigma_1$, effectively classifying membership status.
The use of median values for threshold selection is based on their robustness against outliers and their accurate representation of central tendencies in data distributions. Medians provide a stable threshold, less influenced by extreme values, ensuring consistent performance across diverse datasets.

\begin{algorithm}[!t]
	\SetAlgoLined
	\scriptsize
	\DontPrintSemicolon
	\KwIn{${}^m\mathbf{D}^{shadow}_s$, $\mathbf{D}^n$, $\mathcal{M}_s$, and  $\mathcal{M}_s'$}
	\KwOut{The selected thresholds: $\sigma_1$, $\sigma_2$, $\sigma_3$}
		\For {$\mathbf{x}$ \textbf{in} ${}^m\mathbf{D}^{shadow}_s$}
		{
			$distance_1 \longleftarrow dis(\mathcal{M}_s'(\mathbf{x}), \mathcal{M}_s(\mathbf{x}))$\\
			$\sigma_1  \longleftarrow median(distance_1)$
		}
		\For{$\mathbf{x}_1$, $\mathbf{x}_2$ \textbf{in} $\mathbf{D}^n$, ${}^m\mathbf{D}^{shadow}_s$}
		{
			$distance_2 \longleftarrow dis(d(\mathcal{M}_s(\mathbf{x}_1), \mathcal{M}_s(\mathbf{x}_2)))$\\
			$\sigma_2 \longleftarrow median(distance_2)$
		}
		\For{$\mathbf{x}_1$, $\mathbf{x}_2$ \textbf{in} $\mathbf{D}^n$, ${}^m\mathbf{D}^{shadow}_s$}
		{
			$distance_3 \longleftarrow dis(\mathcal{M}_s(\mathbf{x}_1), |M_s(\mathbf{x}_2)-\mathcal{M}_s'(\mathbf{x}_2)|  )$\\
			$\sigma_3 \longleftarrow median(distance_3)$
		}
		\Return{ $\sigma_1$, $\sigma_2$, $\sigma_3$ }
	\caption{Threshold selection}
	\label{alg:ss}
\end{algorithm}

For establishing the threshold $\sigma_2$, the attacker synthesizes random noisy images to query the student model, subsequently measuring the distances between these queried results and the feature representations from the shadow model's member data. The median of these calculated distances is selected as $\sigma_2$. To set $\sigma_3$, the attacker compares the feature representations obtained from the noisy image queries against the differences found between feature representations from the student model and shadow model member data. The median value from these comparative distances is then assigned as the threshold $\sigma_3$, aiding in the nuanced differentiation of membership inferences.

\section{Other Two Typical Attacks}
\label{sec:othertwo}
We also present two typical MIAs in transfer learning for comparing, where MIAs are considered in a white-box way. 

\textbf{Attack Scheme of At.T \& Ac.T}
As shown in Figure~\ref{fig:ovewflow2}(a), the attacker queries the shadow teacher model using the data that participated in the shadow model training.
The representation of the intermediate layer, $\mathcal{M}_t'(\mathbf{x})$, for $\mathbf{x}\in {}^m\mathbf{D}^{shadow}_t$ is labeled with 1, 
indicating membership. Similarly, the shadow teacher model is also queried using data that did not appear in the shadow model training. 
The output of the intermediate layer, $\mathcal{M}_t'(\mathbf{x})$, for $\mathbf{x} \notin \mathbf{D}_{shadow}$ is labeled with 0, indicating non-membership.
The attacker then uses the output of the intermediate layer and the corresponding labels to train the attack model.
The goal of the attack model is to distinguish the membership status of the shadow teacher model.
In the attack stage, the attacker queries the shadow teacher model using data that participate in the shadow model training.
The output of the intermediate layer, $\mathcal{M}_t'(\mathbf{x})$, for $\mathbf{x}\in {}^m\mathbf{D}^{shadow}_t$ is labeled with 1, indicating membership. 
Similarly, the shadow teacher model is also queried using data that did not appear in the shadow model training. 
The output of the intermediate layer, $\mathcal{M}_t'(\mathbf{x})$, for $\mathbf{x} \notin {}^m\mathbf{D}^{shadow}_t$ is labeled with 0, indicating non-membership.
The attacker then uses the output of the intermediate layer and the corresponding labels, i.e., $(\mathcal{M}_t'(\mathbf{x}), 1)$ and $(\mathcal{M}_t'(\mathbf{x}), 0)$, 
to train the attack model.
The goal of the attack model is to distinguish the membership status of the shadow teacher model.

\textbf{Attack Scheme of At.S \& Ac.S}
As shown in Figure~\ref{fig:ovewflow2}(b), the attacker queries the shadow model using the data that participated in shadow model training. The output of the intermediate layer, $\mathcal{M}_s'(\mathbf{x}), \mathbf{x}\in {}^m\mathbf{D}^{shadow}_s$, is labeled with 1, i.e., belonging to the member.
Similarly, the shadow model is also queried using the data not appearing in shadow model training.
The output of the intermediate layer, $\mathcal{M}_s'(\mathbf{x}), \mathbf{x} \notin {}^m\mathbf{D}^{shadow}_s$, is labeled with 0, i.e., not belonging to the member.
Then the attacker uses the output of the intermediate layer and the corresponding labels,
i.e., $(\mathcal{M}_s'(\mathbf{x}), 1), \mathbf{x}\in {}^m\mathbf{D}^{shadow}_s$ and $(\mathcal{M}_s'(\mathbf{x}), 0), \mathbf{x} \notin {}^m\mathbf{D}^{shadow}_s$, to train the binary classification model.
The model is trained to distinguish the members of the data from the outputs of the intermediate layer.
Similarly, in the attack stage, the adversary queries the student model using the target data record and get the feature representation of hidden layer.
Then, the attacker queries the attack model using the feature representation and obtains the predicted membership status.

\section{Experimental Results}
\label{sec:exp}
In this section, we report performance results.

\subsection{Evaluation Setup}

\textbf{Datasets.}
We used the following widely-used datasets in previous MIA works for evaluation~\cite{ying2020privacy,zou2020privacy,hidano2021transmia,hu2023defenses,nasr2018machine}:
\begin{itemize}[leftmargin=*]
	\item \emph{ImageNet}~\cite{deng2009imagenet} serves as a foundational dataset in computer vision, featuring extensive classes and images, commonly used to train teacher models in transfer learning. The pre-trained models are then adapted to student models for tasks in similar domains, leveraging its detailed class structure.

	\item \emph{CIFAR-100}~\cite{krizhevsky2009learning} comprises 60,000 three-channel color images across 100 varied classes like flowers and fish, each with 500 training and 100 testing images of 32$\times$32 pixels.

	\item \emph{Flowers102}~\cite{102flower} features 102 flower types, totaling 7,169 images distributed across 102 classes with an imbalanced dataset ranging from 40 to 258 images per class. The dataset includes 6,149 training and 1,020 testing images.

\item \emph{Cats vs. Dogs}~\cite{dogcats} dataset includes 25,000 training and 12,500 testing images, evenly divided between cat and dog.

\end{itemize}

\textbf{Experimental Settings.}
ImageNet was used as the teacher dataset to pre-train as teacher model as it contains various classes and has the largest number of images.
During our evaluation, we trained the teacher model using ResNet50, VGG19, Inception v3, and DenseNet169 respectively.
We reimplemented the CNN architectures used in our experiments under TensorFlow 1.15.2.

\begin{figure}[!t]
	\centering
	\includegraphics[width=\linewidth]{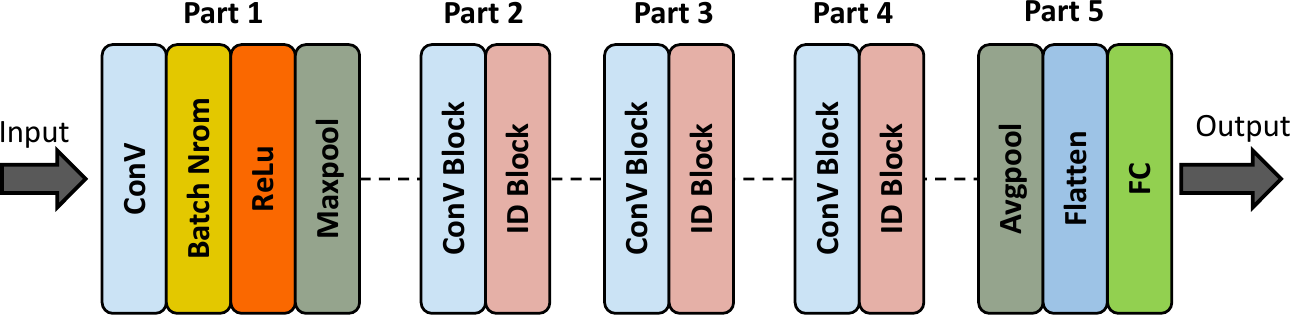}
	\caption{The structure of ResNet50, we divided the model into five parts for transfer learning in evaluation.}
	\label{fig:resnet50}
	\vspace{-4mm}
\end{figure}

In Figure~\ref{fig:resnet50}, we illustrate the widely-used ResNet50 model to evaluate performance of MIAs. The model is segmented into five key parts: Part 1 includes initial convolution, normalization, ReLU activation, and max pooling layers; Parts 2 to 4 contain progressively complex Identity (ID) Blocks for feature extraction; and Part 5 concludes with average pooling and a fully connected (FC) layer for output generation. In our experiments, Parts 1-3 of the teacher model were immobilized to initialize the student models. We fine-tuned these student models using the CIFAR-100, Flowers102, and Cats vs. Dogs datasets, each inheriting pre-trained weights from the teacher model. Additionally, we evaluated various model architectures, including VGG19, InceptionV3, and DenseNet169, experimenting with different freezing configurations (e.g., Parts 1-2, 1-3, and 1-4) to observe the impact on transfer learning effectiveness and subsequent inference attack accuracy.

To ensure the evaluation fairness, we evenly divided the teacher/student datasets into member and non-member groups, ensuring no overlap and a balanced 1:1 ratio for unbiased analysis. Given that the teacher and student datasets originate from distinct sources, their different distributions enhance the assessment's robustness. For constructing shadow models, we split their corresponding datasets into 70\% for training and 30\% for testing, tailoring unique models to each student task. Evaluations were conducted by querying the models with a balanced mix of 10\% member data and an equal amount of non-member data from the testing sets. This process was repeated across ten iterations for each attack scenario to derive consistent performance metrics.


\textbf{Metrics.}
MIA can be perceived as a binary classification challenge: deciding whether a given data point belongs to the target model's training set (positive) or not (negative).
We used the widely-used metrics, including precision, recall, accuracy, and the area under receiver operating characteristic curve (AUC).
Recall is the ratio of accurately classified member data to total member data. 
Precision is the ratio of correctly classified member data to all predicted member data. Accuracy is the ratio of correctly classified data points to all data points. AUC measures the likelihood that the  prediction  score of member data exceed non-member data.

Our evaluations were conducted on Tesla P4*3 GPUs using TensorFlow 2.5.0, with each model trained for 100 epochs, a learning rate of 0.001, and a batch size of 32, utilizing Adam as the optimizer. 
We partitioned the initial dataset, allocating 70\% for training the teacher model and 30\% for the student model, predominantly using the ResNet50 architecture.

\subsection{Overall Performance of At.T \& Ac.S.}
We evaluated the performance of the proposed At.T \& Ac.S of using the input feature representation to perform MIA.
The teacher model was trained using ImageNet dataset and ResNet50 as the base teacher model.
We transferred the teacher model to the student model using CIFAR-100, Flowers102, and Cats vs Dogs dataset, respectively.
Specifically, we froze part 1-3 (Figure~\ref{fig:resnet50}) to train the student model when performing transfer learning.
To perform the attack, the shadow student model was trained from scratch using the shadow student dataset,
where the hidden layer did not contain the behavioral characters of the teacher model.

\begin{figure}[!t]
	\centering
	\begin{minipage}[t]{0.48\linewidth}
		\includegraphics[width=\linewidth]{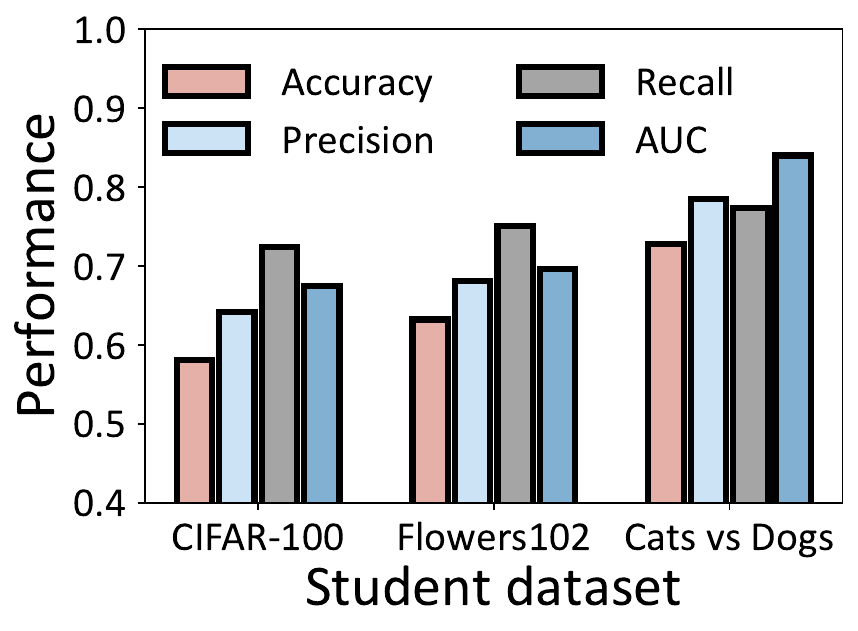}
		\caption{Performance of At.T \& Ac.S under different datasets}
		\label{fig:attacs_performance}
	\end{minipage}
	\hspace{1mm}
	\begin{minipage}[t]{0.48\linewidth}
		\includegraphics[width=\linewidth]{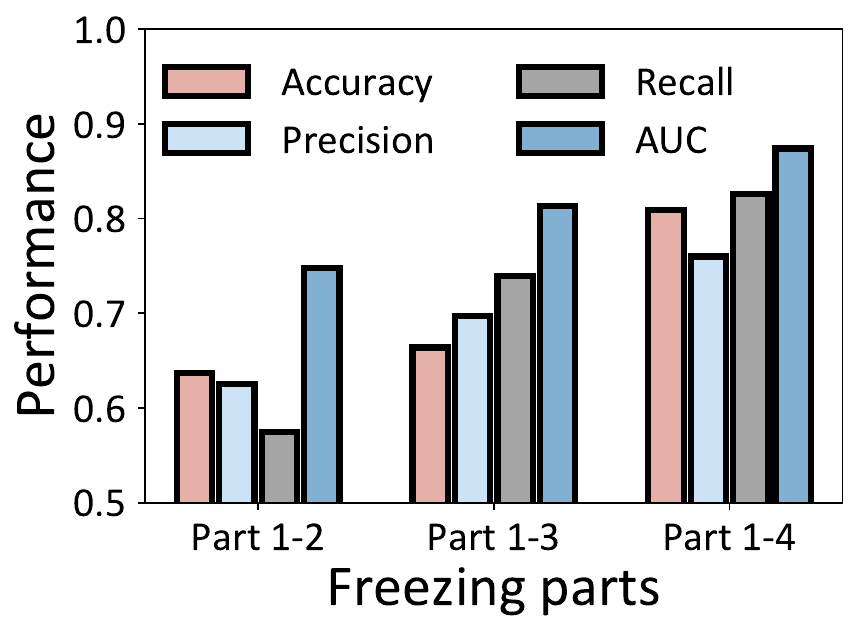}
		\caption{Performance under varied  freezing parts}
		\label{fig:freezingparts}
	\end{minipage}	
\end{figure}

Figure~\ref{fig:attacs_performance} presents the evaluation results of At.T \& Ac.S on three student datasets: CIFAR-100, Flowers102, and Cats vs Dogs. Our method achieves accuracies of 0.581, 0.632, and 0.728, respectively, with AUC values approaching or exceeding 0.7, indicating effectiveness beyond random guessing. The proposed MIA remains effective across most teacher model classes in transfer learning. Notably, the Cats vs Dogs dataset demonstrates higher attack accuracy (0.728) and precision (0.785) compared to CIFAR-100 (0.581, 0.642) and Flowers102 (0.632, 0.681). We speculate this is due to Cats vs Dogs having fewer classes and data points, resulting in smaller changes during transfer learning.


\subsection{Impact of the Number of Frozen Layers}
To evaluate attack performance, we varied the number of frozen layers in the student model, which was trained by freezing different parts of the ResNet50 teacher model, initially trained on the ImageNet dataset. Specifically, ResNet50 was divided into five parts (Figure~\ref{fig:resnet50}), and the student model was trained by freezing parts 1-2, 1-3, and 1-4 respectively. We transferred the model trained on the ImageNet dataset to classify the Cats vs Dogs dataset. Figure~\ref{fig:freezingparts} shows that MIA accuracy improves with more layers frozen: 0.637 for parts 1-2, 0.664 for parts 1-3, and 0.809 for parts 1-4. We speculate that freezing more layers preserves more abstract features, allowing the attacker to better distinguish data records from the teacher model's dataset. Freezing more layers during transfer learning typically preserves pre-learned features, potentially enhancing task accuracy for similar tasks, but may limit adaptability for tasks with different nuances.

Additionally, we also experimented with the CIFAR-100 and Flowers102 datasets. For the CIFAR-100 dataset, we observed a consistent trend with the Cats vs Dogs dataset, where attack performance improved as more layers were frozen, with accuracy varying from 0.527 to 0.681 across different frozen parts. In the Flowers102 dataset, the results were less pronounced but still indicated better attack performance with more frozen layers, with accuracy varying from 0.582 to 0.652. These observations suggest that the relationship between the number of frozen layers and attack efficacy is generally applicable, though the intensity of the effect may vary depending on the dataset characteristics.


\subsection{Comparison with SOTA Methods}
We also conducted a systematic comparison of our proposed MIA with SOTA methods, including Zou et al.~\cite{zou2020privacy} and TransMIA~\cite{hidano2021transmia}.
Zou et al.~\cite{zou2020privacy} conducted an empirical study of MIAs against transfer learning using shadow training techniques,
but did not consider the interconnectedness between the teacher and student models.
TransMIA~\cite{hidano2021transmia} proposed a transfer shadow training method for implementing MIAs against transfer learning, 
where the attacker creates the shadow model by leveraging the parameters of the transferred model.
Specifically, we compared the performance of using intermediate feature representations to perform MIAs.
Typically, we froze first three layers (Figure~\ref{fig:resnet50}) when performing transfer learning to the student model, and the shadow student model was trained from scratch using the shadow student dataset.

\begin{figure*}[!t]
	\centering
	\captionsetup[subfloat]{labelfont=scriptsize,textfont=scriptsize}
	\subfloat[Accuracy]{\includegraphics[width = 0.24\linewidth]{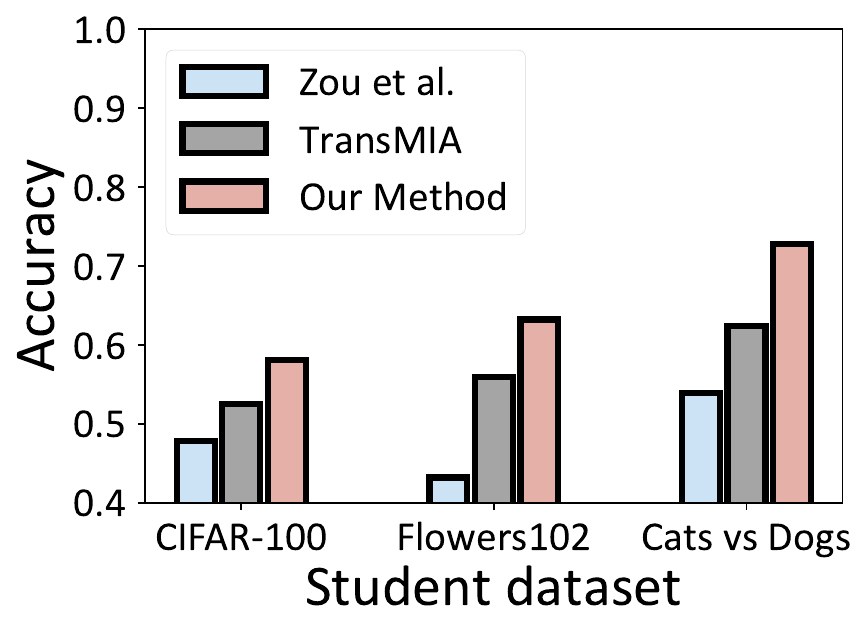}}\hspace{-1mm}
	\subfloat[Precision]{\includegraphics[width = 0.24\linewidth]{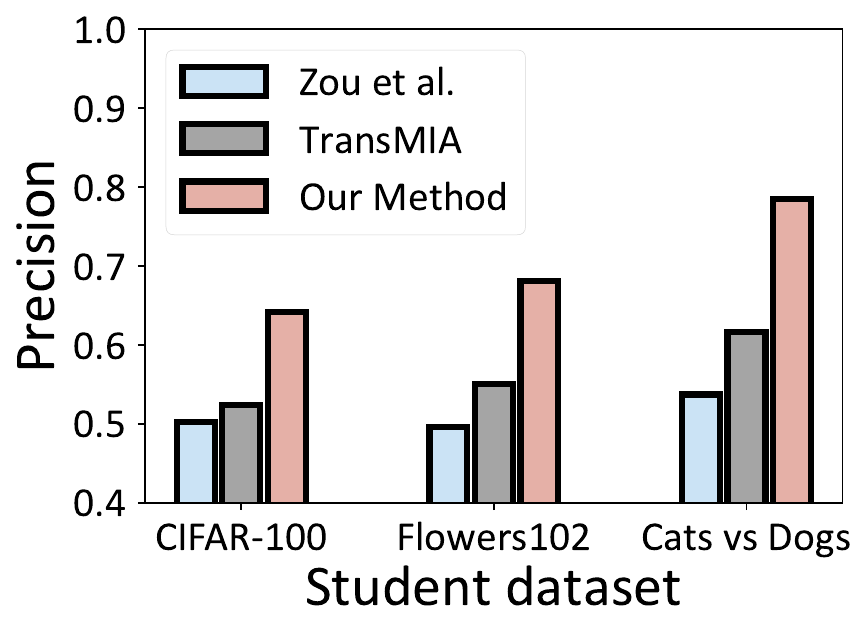}}  \hspace{-1mm}
	\subfloat[Recall]{\includegraphics[width = 0.24\linewidth]{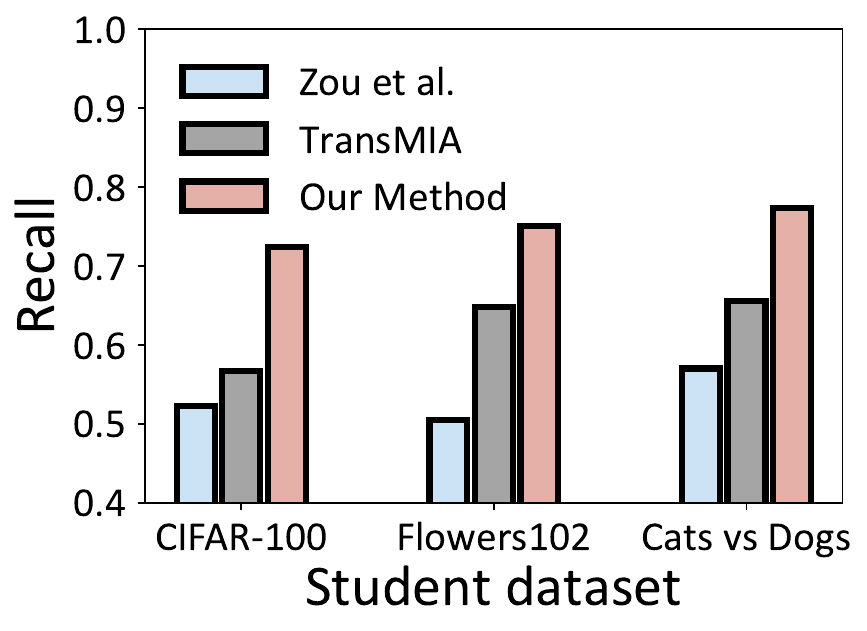}} \hspace{-1mm}
	\subfloat[AUC]{\includegraphics[width = 0.24\linewidth]{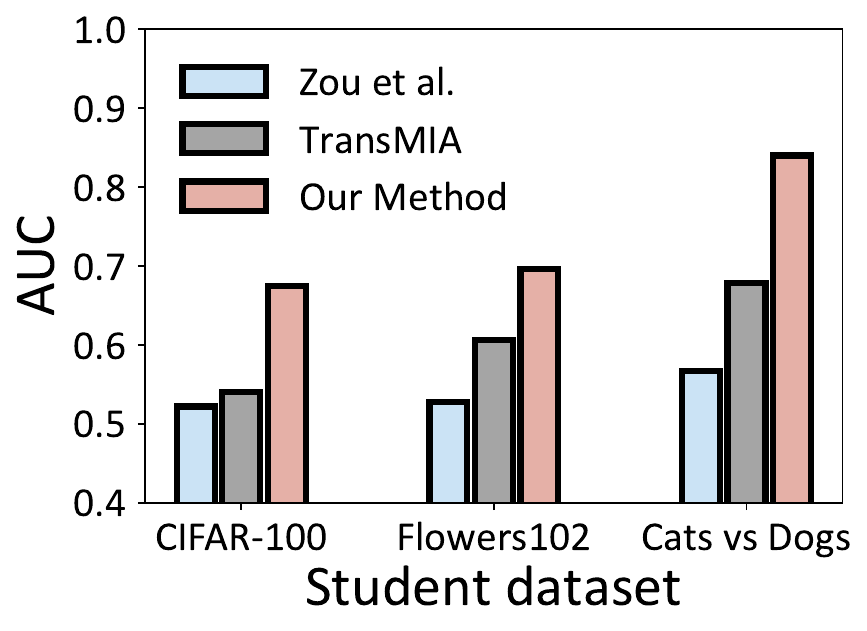}} 
	\vspace{-3mm}
	\caption{ Comparison of SOTA methods and our method on MIA accuracy (a), precision  (b), recall  (c), and AUC  (d).}
	\label{fig:comp_performance}
	\vspace{-3mm}
\end{figure*}

As shown in Figure~\ref{fig:comp_performance}, the performance of our proposed approach is compared to SOTA methods on three datasets. 
The results indicate that our method outperforms existing methods on all three datasets.
For instance, on the student classification task of the cats vs dogs dataset,
our method achieves an accuracy of 0.728, which is superior to the accuracy of 0.624 of TransMIA and 0.539 of Zou et al.
On the student classification task of the CIFAR-100 dataset, our method achieves an accuracy of 0.581, which surpasses the accuracy of 0.525 of TransMIA and 0.478 of Zou et al.
Overall, our proposed MIA is able to expose more membership privacy of the teacher model than SOTA methods.


\subsection{Impact of Different Teacher Models}
We evaluated the impact of different teacher models on the performance of our proposed MIA using four different teacher models.
Specifically, we used VGG19~\cite{he2016deep}, ResNet50~\cite{res50},
Inception v3~\cite{szegedy2016rethinking}, and DenseNet169~\cite{huang2017densely} as the teacher model structure respectively,
since these models achieve the impressive performance and widely used for teacher model training~\cite{wu2020caiauth}.
These models were trained on the ImageNet dataset and transferred to three student datasets.
Note that, we used the same student model structure as the teacher model structure.	
When transferring the teacher model to the student model, we froze the first half of the model so as not to participate in the model training, 
while the back half of the model is included in the model updating.
Specifically, we froze the first three blocks for VGG1, part 1-3 for ResNet50, module \texttt{3$\times$Inception} and \texttt{5$\times$Inception} for Inception v3,and the first two dense blocks for DenseNet169.

Table~\ref{tab:teacher_models} presents the accuracy, precision, recall, and AUC under four teacher models, i.e., VGG16, ResNet50, Inception-v3, and DenseNet169,
and three datasets, i.e., CIFAR-100, Flowers102, and Cats vs Dogs.
The results indicate that our proposed method can be generalized to different teacher models.
For instance, when transferring to the CIFAR-100 dataset, ResNet50, VGG19, Inception v3, and DenseNet169 achieve an accuracy of 0.581, 0.545, 0.568, and 0.593, respectively.
Similarly, when transferring to the Cats vs Dogs dataset, ResNet50, VGG19, Inception v3, and DenseNet169 achieve an accuracy of 0.728, 0.627, 0.700, and 0.725, respectively.
Additionally, it can be observed that VGG is less susceptible to membership inference compared to other model structures in transfer learning.
One possible explanation for this phenomenon may be that VGG is a model with more parameters than the other models,
and the parameters of VGG are more challenging to train using the same setting.
\begin{table}[!t]
	\centering
	\small
	\caption{Performance under different teacher-student model}
	\resizebox{\linewidth}{!}{\begin{tabular}{p{110pt}cccr}
		\hline
		\hline
		Base model   & Accuracy & Precision & Recall & AUC   \\
		\hline
		\hline
		ResNet50 - CIFAR-100   & 0.581  & 0.642     & 0.724  & 0.675 \\
		ResNet50 -  Flowers102  & 0.632  & 0.681     & 0.751  & 0.696 \\
		ResNet50 - Cats vs Dogs   & 0.728  & 0.785     & 0.773  & 0.840 \\
		VGG19  - CIFAR-100    &0.545&0.543&0.572&0.580 \\
		VGG19- Flowers102   &0.587&0.591&0.566&0.631 \\
		VGG19 -Cats vs Dogs    & 0.627&0.632&0.610&0.684 \\
		Inception v3 - CIFAR-100 & 0.568&0.557&0.669&0.608    \\
		Inception v3 - Flowers102& 0.633&0.621&0.681&0.704   \\
		Inception v3 - Cats vs Dogs & 0.700&0.707&0.684&0.774     \\
		DenseNet169  - CIFAR-100& 0.593&0.584&0.650&0.659 \\
		DenseNet169 - Flowers102  &0.671&0.671&0.669&0.739 \\
		DenseNet169 - Cats vs Dogs  &0.725&0.731&0.711&0.801 \\
		\hline
	\end{tabular}}
	\vspace{-4mm}
	\label{tab:teacher_models}
\end{table}


\subsection{Performance of At.T \& Ac.T and At.S \& Ac.S.}
\textbf{Evaluation of At.T \& Ac.T}
As mentioned before, in this attack, the teacher model can be accessed by the adversary,
and the goal is to decide whether the input data record is used to train the student model, 
which is similar to a typical MIA.
We evaluated the performance of At.T \& Ac.T using ImageNet as the teacher dataset under four typical models, 
including VGG19, ResNet50, Inception v3, and DenseNet169.

Table~\ref{tab:performance_of_att} shows accuracy, precision, recall, and AUC of At.T \& Ac.T 
under four teacher models, i.e., VGG16, ResNet50, Inception-v3 and DenseNet169,
The results show that our method achieves an attack accuracy of  0.732, 0.652, 0.751, and 0.769 under the four teacher models, respectively. 
The attack AUC is higher than 0.7, e.g., 0.869, 0.714, 0.830, and 0.849.
Besides, it can be observed that the attack accuracy on ResNet50, Inception v3, and DenseNet169 is higher than VGG19, 
where VGG19 is only with a MIA accuracy of 0.652.

In addition, we speculate that attack accuracies can be significantly affected by the dataset overlap; for example, a teacher model trained on the broad ImageNet and transferred to the specialized Flowers102 might exhibit higher attack accuracies due to the pronounced focus and enhanced feature learning on flower classes, underscoring the role of data relevance and specificity in the transfer learning context for membership inference success.
\begin{table}[!t]
	\centering
	\caption{Attack performance of At.T \& Ac.T under four different model structures}
	\begin{tabular}{lcccc}
		\hline
		\hline
		Base model   & Accuracy & Precision & Recall & AUC   \\
		\hline
		\hline
		ResNet50     & 0.732  & 0.773     & 0.864  & 0.869 \\
		VGG19     & 0.652&0.649&0.665&0.714 \\
		Inception v3 & 0.751&0.766&0.725&0.830   \\
		DenseNet169  & 0.769&0.794&0.729&0.849 \\
		\hline
	\end{tabular}
	\vspace{-3mm}
	\label{tab:performance_of_att}
\end{table}

\begin{figure}[!t]
	\centering
	\includegraphics[width =0.5 \linewidth]{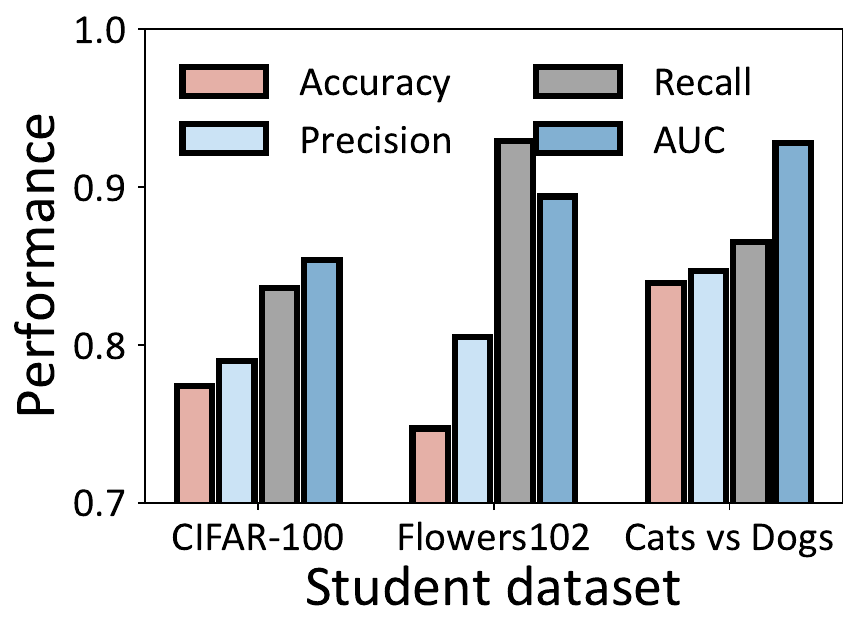}
	\caption{Attack performance of At.S \& Ac.S under CIFAR-100, Flowers102, and Cats vs Dogs dataset.}
	\vspace{-2mm}
	\label{fig:diff-stu}
\end{figure}

\textbf{Evaluation of At.S \& Ac.S.}
We evaluated the performance of At.S \& Ac.S using three student datasets, 
including CIFAR-100, Flowers102, and Cats vs Dogs, 
under the teacher model of ResNet50 trained using ImageNet.
We transferred the teacher model to the three different tasks, respectively.
Specifically, part 1-3 (Figure~\ref{fig:resnet50}) of the teacher model was frozen to train the student model on the student dataset.
Figure~\ref{fig:diff-stu} shows accuracy, precision, recall, and AUC of At.S \& Ac.S under different student datasets.
The results show that our method achieves a high attack accuracy of  0.774, 0.747, and 0.839 on CIFAR-100, Flowers102, and Cats vs Dogs, respectively.
The overall attack AUC is higher than 0.85, e.g., 0.854, 0.894 and 0.928 on three datasets.
We also observed that the attack performance on Cats vs Dogs dataset is higher than CIFAR-100, Flowers102.
We speculated that the reason is that the classification task on Cats vs Dogs is more simple than the other two datasets, 
and thus has a higher classification accuracy when transferring the teacher model to the student task.

\section{Related Work}
\label{sec:related_work}

\subsection{Memebership Inference Attack}
MIAs have been the subject of a significant amount of research, and it has been shown that they can compromise the privacy of the training dataset of the target model~\cite{carlini2022membership,chen2021enhanced,chen2020practical,yeom2018privacy,song2021systematic}.
Existing MIAs can be roughly divided into two categories: classifier-based and metric-based methods.

\textbf{Classifier-based Methods.}
Classifier-based methods in MIAs use binary classifiers trained through shadow training to mimic the target model's behavior and predict data point membership status, effectively utilizing training and testing data to generate informative datasets for inference~\cite{shokri2017membership,salem2018ml,long2020pragmatic}. Shokri et al.~\cite{shokri2017membership} were pioneers, employing shadow training for MIAs with black-box access, setting a foundational methodology for others like Salem et al.~\cite{salem2018ml}, who optimized this approach to enhance attack accuracy. Meanwhile, Long et al.~\cite{long2020pragmatic} introduced strategic targeting within MIAs, underscoring the strategic nuance in identifying and exploiting model vulnerabilities.


\textbf{Metric-based Methods.}
Metric-based methods in MIAs rely on analyzing prediction vector metrics against specific thresholds to ascertain membership status~\cite{yeom2018privacy,song2021systematic}. Yeom et al. highlighted a novel MIA approach focusing on prediction accuracy and loss, indicating that risks extend beyond model overfitting~\cite{yeom2018privacy}. Song et al. critiqued existing entropy-based MIAs for neglecting true label characteristics, proposing a refined method integrating prediction entropy with the actual label for more accurate inference~\cite{song2021systematic}. Similarly, Hui et al. introduced a shadow-less blind membership inference utilizing prediction entropy to gauge data's training involvement without shadow models~\cite{hui2021practical}. Li et al. embraced a black-box framework, leveraging adversarial perturbation to maintain efficacy even with limited model output~\cite{li2021membership}, while Liu et al. adopted a distilled loss trajectory approach for membership detection, analyzing losses across model training stages~\cite{liu2022membership}.

Recent research has examined white-box MIA scenarios, where attackers obtain comprehensive access to the target machine learning model's architecture and parameters. Nasr et al.~\cite{nasr2019comprehensive} discovered that while final predictions and intermediate computations alone did not surpass black-box attack accuracies, integrating gradients of the prediction loss relative to model parameters significantly improved attack effectiveness in white-box contexts. Similarly, Jayaraman et al.~\cite{jayaraman2021revisiting} proposed an MIA that leveraged the loss variability caused by perturbations, showing its effectiveness in datasets with class imbalances.
Besides, MIAs have been documented in adjacent domains like P2P federated learning and clustered federated learning, where transfer learning features are utilized. For instance, Luqman et al.~\cite{luqman2023membership} demonstrated that federated learning structures are susceptible to MIAs due to the shared learning frameworks.
	
\subsection{MIAs Against Transfer Learning}
Transfer learning, crucial for both industrial and academic progress~\cite{wu2022echohand}, faces significant security risks, including data poisoning~\cite{schuster2020humpty}, backdoor attacks~\cite{yao2019latent}, and susceptibility to adversarial examples~\cite{wang2018great}. Within this context, the exploration of MIAs by Zou et al.\cite{zou2020privacy} and Hidano et al.\cite{hidano2021transmia} has advanced understanding of these vulnerabilities, though limitations in attack accuracy remain due to insufficient exploitation of complex teacher-student model interrelations.

Zou et al.\cite{zou2020privacy} conducted an empirical investigation into MIAs under transfer learning, facing challenges in achieving high accuracy in black-box attacks, as shown in their CIFAR-100 dataset results. Hidano et al.\cite{hidano2021transmia} improved attack effectiveness using a white-box approach with a transfer shadow training strategy, but this required specific shadow model configurations of the teacher dataset. Our research builds on these studies, adopting a nuanced white-box framework that thoroughly examines differential feature representations between teacher and student models. By dissecting the intricacies of knowledge transmission and representation, our approach aims to enhance the precision of membership inference, expanding the discussion on securing transfer learning frameworks against such attacks.

Our approach leverages the interplay between teacher and student models, analyzing hidden layer representation discrepancies to infer membership status. Unlike conventional black-box methodologies that may overlook this interconnection, our white-box-centric method, adaptable to black-box scenarios via surrogate modeling, delves deeper into model internals for more insightful analysis. Prior studies, such as Zou et al.\cite{zou2020privacy}, may not fully address teacher-student model interconnectivity due to black-box constraints. In contrast, our method enhances attack efficacy in real-world transfer learning contexts by capitalizing on this aspect. Additionally, we advance beyond TransMia\cite{hidano2021transmia} by integrating insights from both teacher and student models, rather than isolating the student model analysis. Table~\ref{tab:comparison} presents a high-level comparison of our approach and existing works.

\begin{table}[!t]
	\centering
	\scriptsize
	\caption{Comparison of our approach and existing works}
	\begin{tabular}{lp{55pt}p{55pt}p{55pt}}
		\hline
		\textbf{Aspect} & \textbf{Our Approach} & \textbf{Zou et al.\cite{zou2020privacy}} & \textbf{TransMia\cite{hidano2021transmia}} \\ \hline
		\textbf{Scenario} & black-box & Black-box & Black-box \\ 
		\textbf{Focus} & Teacher-student model & General privacy attacks & Student model \\ 
		\textbf{Methodology} & Hidden layer representation analysis & Shadow model training & Only student model analysis \\ \hline
	\end{tabular}
	\label{tab:comparison}
\end{table}

\section{Discussion}
\label{sec:dis}
This section discusses mitigation strategies and limitations.

\subsection{Attack Mitigation Strategies}
As of now, specific defense mechanisms to mitigate MIAs in transfer learning are not well-defined~\cite{hu2023defenses}, limiting our ability to compare our study against established MIA defenses and highlighting an urgent research gap. Future investigations are imperative to devise and assess tailored defenses that can effectively protect privacy within transfer learning contexts. 

In this light, we outline potential defense strategies to address MIAs in transfer learning environments:
\emph{(i) Output randomization and differential privacy}, applying noise addition to model parameters and data or using subset-based training and prediction thresholding~\cite{backes2016membership}, while it may reduce the model performance; 
\emph{(ii) Adversarial training}, integrating adversarial examples into training to fortify defenses~\cite{jia2019memguard}; \emph{(iii) Generative learning approaches}, leveraging generative adversarial networks to obscure membership signals; 
\emph{(iv) Model splitting}, distributing training across various models on disjoint data subsets to obfuscate membership clues; 
and \emph{(v) Model pruning}, trimming unnecessary parameters to minimize information leakage and potentially pruning the teacher model pre-transfer to attenuate the linkage with the student model, thereby enhancing privacy protections.

\subsection{Limitation and Future Work}

Despite our diligent efforts to maintain the validity of our study, certain limitations remain. We focused solely on image classification tasks, omitting other domains such as natural language processing, medical analysis, and object detection. Additionally, our attack method targets typical parameter-based transfer learning, excluding other types such as instance-based, feature representation-based, and relational knowledge-based transfer learning~\cite{pan2010survey, lin2013double, jeng2020low, wang2016relational}. Evaluating membership inference attacks across various transfer learning methods and tasks is crucial. Our study mainly considers attacks on the teacher model with prior knowledge from the student model. Future research should explore attack performance when examining prediction vectors for different classes. While we used Euclidean distance as the primary metric, other metrics like Manhattan and Cosine distances showed lower MIA accuracy; thus, investigating alternative metrics, such as Mahalanobis distance, is essential. We observed that freezing larger parts of the model increases recall and performance, which requires further mathematical substantiation in future work.

\textbf{Assumptions.} While our model assumes white-box access, offering a stringent evaluation context, the comparison with existing works~\cite{zou2020privacy,tian2023manipulating,hu2023defenses}'s black-box scenario is to illustrate how model accessibility influences attack success. Acknowledging the potential difference in real-world shadow and teacher/student dataset distributions, our experiment uses the same dataset to underscore our approach's effectiveness under less-than-ideal conditions. Future research will assess this in more varied scenarios.

Our attacker model, grounded in white-box assumptions, reflects practical adversarial capabilities in contemporary cybersecurity landscapes, where attackers can access or infer detailed model information. This model's practicality is substantiated by real-world scenarios wherein entities might gain insider information or exploit system vulnerabilities to access model details, elevating our approach beyond traditional black-box methods. By delving into the nuanced interconnectivity between teacher and student models, our methodology not only showcases superior performance over existing black-box approaches like~\cite{zou2020privacy,hidano2021transmia} but also demonstrates a pragmatic understanding of attack vectors, thereby enhancing the relevance and applicability of our findings in addressing and mitigating real-world transfer learning vulnerabilities.

In addition, while attack mitigation strategies can reduce attack efficacy, it is important to consider their impact on model performance. Future research should focus on balancing privacy protection with accurate predictions, especially in sensitive applications where membership inference attacks pose significant risks.
Also, future work will involve a detailed empirical investigation into the dimensionality and feature characteristics of different datasets to better understand their impact on attack performance.

To adapt our methodology to the black-box setting, attackers can use iterative queries to approximate the target model's behavior through a shadow model, which mimics the target model's responses. By analyzing discrepancies in output probabilities and refining the shadow model, attackers can identify membership patterns via probabilistic analysis. This adaptation necessitates careful threshold tuning to maximize inference accuracy without direct access to internal model parameters. In a typical teacher → student → student transfer scenario, each model retains features from the previous one, preserving patterns indicative of the original teacher's training data. Our methodology can effectively trace these patterns, even across multiple transfer levels. Future research should investigate the resilience of this attack approach in complex transfer chains to strengthen privacy defenses.

\textbf{Model Structure.}
Our evaluation utilizes popular transfer learning models, including ResNet50, VGG19, Inception v3, and DenseNet169~\cite{wu2021toward,wu2022echohand}, selected for their optimal balance of computational efficiency and robust performance in real-world applications. Their depth and complexity enable nuanced analysis of hidden layer representations between teacher and student models. By leveraging deep learning principles of feature representation and transferability, the method is robust and generalizable across architectures with hierarchical feature processing. Preliminary tests using ResNet50 have yielded promising results, underscoring the method's potential. To confirm its broad applicability and adaptability, future research should evaluate additional models, e.g., Inception v2 and EfficientNet, and different frameworks, e.g., transformer across varying architectures and complexities.

\section{Conclusion}
\label{sec:conclusion}
In this study, we introduce a new MIA methodology in transfer learning, elucidating the teacher model's privacy vulnerabilities. By analyzing the nuanced interplay between teacher and student models, our approach effectively identifies and leverages differences in feature representations to infer the knowledge transferred. Our extensive evaluations confirm the efficacy of this method, revealing its capability to uncover more significant privacy details from the teacher model compared to existing SOTA approaches. These findings underscore the critical need to account for the dynamic relationship between teacher and student models when assessing and fortifying against membership inference threats in transfer learning scenarios. As edge computing evolves, it offers a promising direction for future research in serverless multi-cloud edge computing, potentially enabling more efficient and secure distributed systems at scale~\cite{duan2025rethinking,liang2024vulseye,wu2024wafbooster,liang2024ponziguard}. As a potential future direction, we are looking forward to extending our method to improve the performance of various applications such as large language models~\cite{lin2023pushing,wu2024semantic,lin2024splitlora,fang2024automated,han2024effectiveness}, distributed learning system~\cite{lin2024adaptsfl,lyu2023optimal,lin2025leo,sun2024efficient,lin2024split}.

\bibliographystyle{IEEEtran}
\bibliography{library}

\end{document}